\documentclass[usenatbib]{mn2e}
\usepackage{epsfig,rotating,natbib,times}

\def\bv{\mathbf{v}}
\def\bA{\mathbf{A}}
\def\bB{\mathbf{B}}
\def\br{\mathbf{r}}

\def\curl{\nabla\times}

\def\smb{\sum_b m_b}

\def\gwab{\nabla_a W_{ab}}
\def\divB{\nabla\cdot\mathbf{B}}
\newcommand{\pder}[2]{\ensuremath{\frac{\partial #1}{\partial #2}}}

\title[Smoothed Particle Magnetohydrodynamics]{Smoothed Particle Magnetohydrodynamics \\ I. Algorithms and tests in one
dimension}
\author[Price \& Monaghan]{D.J. Price$^1$, J.J. Monaghan$^2$ \\
$^1$Institute of Astronomy, Madingley Rd, Cambridge, CB3 0HA, UK \\
$^2$School of Mathematical Sciences, Monash University, Clayton 3800, Australia\\
}
\date{Submitted: 6th June 2003 Accepted: 28th Oct 2003}
\pagerange{\pageref{firstpage}--\pageref{lastpage}} \pubyear{2003}

\begin{document}
\label{firstpage}
\bibliographystyle{mn2e}
\maketitle

\begin{abstract}
 In this paper we show how the Smoothed Particle Hydrodynamics (SPH) equations
for ideal magnetohydrodynamics (MHD) can be written in conservation form with
the positivity of the dissipation guaranteed. We call the resulting algorithm
Smoothed Particle Magnetohydrodynamics (SPMHD). The equations appear to be
accurate, robust and easy to apply and do not suffer from the instabilities known to exist
previously in formulations of the SPMHD equations. In addition we formulate our
MHD equations such that errors associated with non-zero divergence of the
magnetic field are naturally propagated by the flow and should therefore remain small.

 In this and a companion paper \citep{pm03b} we present a wide range of
numerical tests in one dimension to show that the algorithm gives very good
results for one dimensional flows in both adiabatic and isothermal MHD. For the one dimensional tests
the field structure is either two or three dimensional.

 The algorithm has many astrophysical applications and is particularly
suited to star formation problems.
\end{abstract}


\section{Introduction}

 Star forming regions are known to contain magnetic fields which are
sufficiently strong to play an important part in the formation of the dense
concentrations of matter which lead to stars. In order to describe the
dynamics of such a system it is customary to begin with the equations of
ideal magnetohydrodynamics (MHD). However,  the simplicity of the MHD equations
is deceptive and hides the fact that there are numerous technical difficulties
involved in their solution. Our aim in this paper is to describe a set of
Smoothed Particle Hydrodynamics (SPH, for a review see \citealt{monaghan92}) 
equations which overcome these difficulties and can be used to simulate MHD
phenomena.  We call the resulting algorithm Smoothed Particle
Magnetohydrodynamics (SPMHD). The equations appear to be
accurate, robust and easy to apply and they do not suffer from the instabilities
that exist in other formulations of the SPMHD equations.

 An early application of SPH to MHD problems was to static magnetic polytropes
\citep{gm77} who found good agreement with perturbation calculations.  Dynamical
problems were considered by \citet{p83a} and applied to star formation problems
\citep{p82,p83b,p85,p86a,p86b,benz84,pm85}. 
In the latter it was shown that when the conservation form of the equations was
used an instability developed which took the form of SPH particles clumping. 
SPH blast waves in a magnetic medium were studied by \citet{sp90,sp94}.
\citet{hea91}, \citet*{mwb96} and \citet{mkbs99}
used a form of the SPH equations where the magnetic fields were updated on a
grid and interpolated to the SPH particles. 

\citet{m94,m95} and \citet{mwd95} used a formulation of SPMHD that uses a non-conservative ($\bf J
\times B$) force, which is always stable and guarantees
that the magnetic force is exactly perpendicular to the magnetic field. This formalism was also used by
\citet{bp96} and more recently by Cerqueira and de Gouveia dal Pino (2001 and
references therein) and \citet{hosking02}, however the non-conservation of momentum leads to poor
performance on shock-type problems. A conservative form of SPMHD has been used by \citet*{dbl99} and
by \citet*{mal01} since the magnetic field in their simulations remained in the regime
where the instability does not appear. \citet{mphd96} suggested using a
compromise between the conservative (tensor) force and the $\bf J
\times B$ formalism. Non-ideal MHD terms in SPH were also considered by \citet{mphd96}, who suggested using
resistive terms to control the divergence of the magnetic field and by
\citet{hosking02}, who considered the effects of ambipolar diffusion via a
two-fluid model.

  The first technical difficulty with MHD simulations is that the magnetic field
comes with the constraint that $\nabla \cdot {\bf B} = 0$.  \citet{bb80}
showed that in some finite difference codes, the failure to satisfy this
constraint would lead to an instability.  A number of different techniques
have therefore been developed to ensure that this constraint is satisfied. 
The first of these is to work with the vector potential $\bf A$ where ${\bf B}
= \nabla \times {\bf A}$ rather than with ${\bf B}$.  This approach was used
by \citet{gm77} for SPH simulations. 
Others \citep{eh88,sn92} construct their finite difference equations so that, to
the accuracy of the resolution, $\nabla \cdot {\bf B}=0$.  Another approach commonly used is
to clean up the magnetic field at every step by adding a gradient term to the
computed field to produce a new field which satisfies the constraint.   By
giving up the conservation of momentum and including $\nabla \cdot {\bf B}$
terms in the momentum equation \citet{powell94} (see \citealt{pea99})
produced a stable finite difference scheme for MHD (the 8 wave theory) which,
however, appears to be less accurate for shocks.  A comprehensive discussion
of these and other schemes has been given by \citet{toth00} who notes that
even if $\nabla \cdot {\bf B} = 0$ most of these schemes produce magnetic
forces which are not perpendicular to the exact magnetic force ${\bf J}\times
{\bf B} $.  Within the framework of SPH \citet{b01} (see \citealt*{bot01})
developed a non-conservative form of the SPMHD equations which have good
stability properties and by the creation of closely spaced particles in
regions with large spatial gradients it gives excellent accuracy.  

An alternative to any of these approaches is that of \citet{janhunen00} who
starts from the premise that non zero $\nabla \cdot {\bf B}$ terms may be
generated but,  if they are treated consistently, no instabilities will occur. 
The resulting set of equations has been derived by \citet{dellar01} starting with
the relativistic formulation of gas dynamics with electromagnetic fields.
\citet{janhunen00} showed that this formulation of MHD gas dynamics could be
simulated using the HLL method \citep*{hll83} and he showed by thousands of test cases that
positivity could be expected even if not proven. 

Our approach is to follow \citet{janhunen00} and use equations which are
consistent even if $\nabla \cdot {\bf B}$ does not vanish.  To simulate shocks
we introduce an artificial dissipation which guarantees that changes to entropy
and thermal energy from viscous and ohmic dissipation are positive. The
resulting set of equations conserves momentum and energy exactly. 

Another technical difficulty peculiar to SPH is that when a conservative force
is used the SPH particles tend to clump in pairs in the presence of tension.  This was first noticed by
\citet{pm85} and re-discovered by researchers applying SPH to elastic
fracture problems (see the references in \citealt{monaghan00}).  Several remedies have
been proposed (e.g. \citealt*{dri97,bk00,bk01}) but
they all either involve a significant increase in computation or cannot be
applied where the particle configuration changes significantly.  A remedy for the
tensile instability which can be easily applied to astrophysical problems was
proposed by \citet{monaghan00}. The idea is to add a small artificial stress which
prevents particles from clumping in the presence of a negative stress.  This
term has been shown to work well in elastic dynamics simulations \citep{gms02}
and we apply it here to the MHD case.  We find that such a term
very effectively removes the instability with few side effects.

 In \S\ref{sec:continuum} we give the continuum form of the equations, and in
\S\ref{sec:sphforms} the SPH form of these equations. We construct appropriate dissipation
terms for MHD in \S\ref{sec:dissipation}. The instability correction is discussed in
\S\ref{sec:instability} with details in the appendix. The time-stepping strategy
is described in \S\ref{sec:timestep}. In \S\ref{sec:1Dtests} we
present the results of extensive numerical tests for one dimensional
problems involving discontinuous initial conditions. In a companion paper
(\citealt{pm03b}, hereafter paper II)
we derive the SPMHD equations from a variational principle, including the case
where the smoothing length is regarded as a function of local particle density.
A self-consistent derivation of the SPMHD equations in the latter case is shown
to increase the accuracy of SPMHD wave propagation. Two and
three dimensional tests will be presented elsewhere.

\section{The continuum equations}
\label{sec:continuum}

In the absence of dissipation the  $ i^{th}$ component of the acceleration equation is
\begin{equation}
\frac{dv^i}{dt}  = \frac{1}{\rho} \frac{\partial S^{ij} }{\partial x^j},
\end{equation}
where $d/dt$ denotes the derivative following the motion, and the stress  $S^{ij}$ in the case of ideal MHD is defined by
\begin{equation}
S^{ij} = -P \delta^{ij}  + \frac{1}{\mu_0} ( B^i B^j- \frac12 \delta^{ij}  B^2 ),
\end{equation}
Here $B^i$ is the $i^{th}$ component of the magnetic field and $\mu_0$ is the
permittivity of free space.  For  SI units $\mu_0 = (4 \pi)/10^7$.

The time change of the magnetic field is given by the induction equation.  We
follow \citet{janhunen00} and \citet{dellar01} and construct the induction
equation so that it is consistent even if $\nabla \cdot {\bf B} $ does not
vanish. The induction equation including ohmic dissipation then becomes 
\begin{equation}
\frac{\partial {\bf B} }{\partial t} +  \nabla \times ({\bf v} \times {\bf B}) =  - \nabla \times (\eta {\bf J})- {\bf v (\nabla \cdot B}),
\label{eq:induction1}
\end{equation}
where the last term is the magnetic current \citep{janhunen00,dellar01} and ${\bf J}$ is the normal current 
\begin{equation}
 {\bf J } = \mu_0 \nabla \times {\bf B}.
\end{equation}
and $\eta$ is the magnetic diffusivity $1/(\sigma  \mu_0)$  where $\sigma$ is the  conductivity. 

This induction equation can be written 
\begin{equation}
\frac{d {\bf B} }{d t} = ({\bf B} \cdot \nabla ){\bf v} -  {\bf B } (\nabla \cdot {\bf v} ) - \nabla \times (\eta {\bf J}).
\label{eq:induction}
\end{equation}
This last form of the induction equation is the standard form when the
constraint $\nabla \cdot {\bf B} = 0$  is used.  Magnetic monopoles associated
with $\nabla \cdot {\bf B} \ne 0$ do not affect this equation. Taking the
divergence of (\ref{eq:induction1}), we find that monopoles evolve according to
\begin{equation}
\pder{}{t} (\divB) + \nabla\cdot (\bv \divB) = 0,
\end{equation}
which has the same form as the continuity equation for the density and therefore
implies that the volume integral of $\divB$ is conserved.

It is common to solve the acceleration equation with the thermal energy and
continuity equations, but in this section we will assume that the thermal energy
equation is replaced by the total energy equation. Our aim is to derive a set
of SPH equations which conserve total energy and momentum while ensuring that
the change in entropy due to dissipation is positive.

The total energy $e$ per unit volume is defined by
\begin{equation}
e = \rho \left ( \frac12 v^2 + u +\frac{ B^2}{\rho \mu_0} \right ),
\end{equation}
where  $u$ is the thermal energy/unit mass.  The equation for the rate of change of $e$ can be written in terms of the stress according to

\begin{equation}
\frac{\partial e}{\partial t} = - \nabla \cdot  (e {\bf v})+ \frac{\partial (v^i S^{ij})}{\partial x^j}  + \nabla \cdot ({\bf B } \times  (\eta {\bf J} )),
\end{equation}
or as 
\begin{equation}
\frac{\partial e}{\partial t} = - \nabla \cdot \left  [ \left (e + P +
\frac{B^2}{2\mu_0} \right ) {\bf v} - \frac{ {\bf B }({\bf v \cdot B}) }{\mu_0}  - {\bf B } \times  (\eta {\bf J} ) \right ].
\end{equation}

To derive SPH equations it  is convenient to  replace the energy/unit volume by the energy per unit mass $\widehat \epsilon$ 

\begin{equation}
 \widehat \epsilon   = \frac12 v^2 + u +\frac{ B^2}{2\rho \mu_0} .
\label{eq:ehat} 
\end{equation}

 The equation for $\widehat \epsilon$ is 
\begin{equation}
\frac{d \widehat \epsilon}{dt} = \frac{1}{\rho} \frac{\partial  (S^{ij} v^j
) }{\partial x^j} + \frac{1}{\rho} \nabla \cdot ({\bf B} \times (\eta {\bf J}) ).
\end{equation}
 
The thermal energy equation can be derived either from (\ref{eq:ehat}) giving

\begin{equation}
\frac{du}{dt} = \frac{d \widehat \epsilon}{dt} - {\bf v} \cdot \frac{d {\bf v} }{dt}  - \frac{d}{dt } \left ( \frac{{B^2}}{2\mu_0 \rho } \right ),
\label{eq:ufrome}
\end{equation}
or by using the first law of thermodynamics including the ohmic heating term.  Either way we find

\begin{equation}
\frac{du}{dt} =  - \frac{P}{\rho} \nabla \cdot {\bf v} +  \eta J^2,
\end{equation}

The final equation is the density equation 

\begin{equation}
\frac{d\rho}{dt} = - \rho \nabla \cdot {\bf v}.
\end{equation}

In addition we need an equation of state and an expression for the
conductivity.  In this paper we assume the gas is an ideal gas and we
introduce an artificial dissipation into the SPH equations which we then
interpret in terms of an artificial viscosity, thermal conductivity and
magnetic diffusivity.
\section{The SPH equations}
\label{sec:sphforms}
We take as our fundamental equations the acceleration equation, the total
energy equation and the density equation. To construct our SPMHD equations we
follow the procedure used for the special relativistic equations \citep{cm97}. 
Initially the equations will be set up assuming there is no ohmic dissipation. 
We will introduce artificial dissipation in the SPMHD equations in order to
handle shocks and we will then show that to guarantee that the ohmic
dissipation is always positive we must include an extra term in the induction
equation. This extra term gives the appropriate extension of the induction
equation to include the effects of an artificial conductivity.

The acceleration equation for SPH particle $a$ is \citep{monaghan92}
\begin{equation}
\frac{dv_a^i}{dt} = \sum_b m_b \left ( \frac{S_a^{ij} }{\rho_a^2}  + \frac{S_b^{ij} }{\rho_b ^2} + \Pi_{ab} 
                  \right ) \frac{\partial W_{ab} }{\partial x_a^j },
\label{eq:tensor}
\end{equation}
where $W_{ab} = W(\vert\br_a - \br_b\vert, h)$ is the smoothing kernel. The dissipation term $\Pi_{ab}$ will be discussed shortly.  We write the energy equation in the absence of ohmic dissipation in the  form
\begin{equation}
\frac{d \widehat \epsilon}{dt} =   v^i  \frac{\partial}{\partial x^j} \left (  \frac{ S^{ij} }{\rho}  \right ) + \frac{S^{ij}}{\rho^2} \frac{\partial ( \rho v^i) }{\partial x^j},
\end{equation}
which is similar to that used for special relativistic SPH \citep{cm97}.  The SPH equivalent of this equation is 
\begin{equation}
\frac{d \widehat \epsilon_a}{dt} = \sum_b m_b \left (\frac{v^i_a S_b^{ij} }{  \rho_b^2}  +  \frac{v^i_b S_a^{ij} }{  \rho_a^2}  + \Omega_{ab}  \right ) \frac{\partial W_{ab}} {\partial x_a^j},
\label{eq:sphenergy}
\end{equation}
where $\Omega_{ab}$ is a dissipation term analogous to $\Pi_{ab}$.  Because of
the symmetry of the terms in the summation total linear momentum $\sum_a m_a
{\bf v}_a$,  and energy $\sum_a m_a\widehat \epsilon_a$ are conserved.

The induction equation  in the absence of the ohmic term can be written   
\begin{equation}
\frac{dB^i}{dt} = \frac {B^j}{\rho}  \frac {\partial }{\partial x^j}  (\rho v^i) - \left (B^j \frac {\partial \rho}{\partial x^j}   \right )  \frac {v^i}{\rho} - \frac {B^i}{\rho} \left  ( \frac {\partial (\rho v^j) } {\partial x^j} - v^j \frac {\partial \rho}{\partial x^j} \right).
\end{equation}
The SPH form of this equation is 
\begin{equation}
\frac{dB_a^i}{dt} = \frac{1}{\rho_a} \sum_b m_b \left ( v_{ba}^i B_a^j  - B_a^i v_{ba}^j  \right ) \frac{\partial W_{ab}}{\partial x_a^j},
\label{eq:sphinduction}
\end{equation}
where $v_{ba}^j $ denotes $(v_b^j - v_a^j)$. Equivalently we can use
\begin{equation}
\frac{d}{dt}\left(\frac{B_a^i}{\rho_a}\right) = \frac{1}{\rho_a^2}\smb v^i_{ba}
B^j_a \frac{\partial W_{ab}}{\partial x_a^j}.
\label{eq:Bevolsph2}
\end{equation}

 As is usual practice in SPH, the density is estimated via a summation over
neighbouring particles according to
\begin{equation}
\rho_a = \smb W_{ab},
\label{eq:rhosum}
\end{equation}
or alternatively using the time derivative of this expression which gives the 
SPH form of the continuity equation
\begin{equation}
\frac{d\rho_a}{dt} = -\smb v^i_{ba} \frac{\partial W_{ab}}{\partial x_a^i}.
\label{eq:sphcty}
\end{equation}

 We note that equations (\ref{eq:tensor}) and (\ref{eq:sphenergy}) can be derived from a variational
principle using (\ref{eq:sphinduction}) and (\ref{eq:sphcty}) as constraints,
demonstrating that these are indeed a consistent set of equations. This is presented
in paper II. 

 The smoothing kernel we use is the usual spline-based kernel, given by
\begin{equation}
W(q) = \frac{\sigma}{h^\nu}\left\{ \begin{array}{ll}
1 - \frac{3}{2}q^2 + \frac{3}{4}q^3, & 0 \le q < 1; \\
\frac{1}{4}(2-q)^3, & 1 \le q < 2; \\
0 & q \ge 2 \end{array} \right.
\end{equation}
where $q = \vert \br_a - \br_b \vert / h$, $\nu$ is the number of spatial
dimensions and the normalisation constant $\sigma$ is given by $2/3$, $10/(7\pi)$ and $1/\pi$ in
1, 2 and 3 dimensions respectively. The smoothing length $h$ of particle $a$ is set according to the usual rule
\begin{equation}
h_a \propto \left(\frac{1}{\rho_a} \right)^{(1/\nu)}.
\label{eq:hrho}
\end{equation}
We implement this by evolving the smoothing length according to
(\citealt{benz90,monaghan92})
\begin{equation}
\frac{dh_a}{dt} = -\frac{h_a}{\nu\rho_a}\frac{d\rho_a}{dt},
\label{eq:hevol}
\end{equation}
 This works extremely well for the tests presented in this paper since the
density is evolved using the continuity equation (\ref{eq:sphcty}). We note,
however, that the
dependence of the smoothing length on the density given by (\ref{eq:hrho}) can
be used to derive (again via a variational principle) a set of discrete
equations for both SPH and SPMHD which
self-consistently account for the extra terms which arise from this dependence.
This set of equations is derived and implemented in paper II, where we demonstrate
that it leads to increased accuracy in the propagation of MHD waves. In particular the
formalism derived in paper II is natural to use when the density is calculated via
the SPH summation (\ref{eq:rhosum}). In this paper we simply take the average of
the kernel to maintain the symmetry in the momentum and energy equations
\citep{hk89,monaghan92}, that is
\begin{equation}
W_{ab} = \frac{1}{2} \left[W_{ab}(h_a) + W_{ab}(h_b)\right],
\end{equation}
and correspondingly
\begin{equation}
\pder{W_{ab}}{x^i_a} = \frac{1}{2} \left[\pder{W_{ab}(h_a)}{x^i_a} +
\pder{W_{ab}(h_b)}{x^i_a} \right].
\end{equation}

\section{Dissipation Terms}
\label{sec:dissipation}
\citet{cm97} discuss the SPH dissipation analogously to that associated with
Riemann solvers. The key point  is that the dissipation involves jumps in
appropriate variables (momentum, energy and density) between the left and right
Riemann states multiplied by eigenvalues which can be interpreted as signal
velocities. In the SPH case we construct dissipation terms in a similar way.
Thus
\begin{equation}
\Pi_{ab} = - \frac { K v_{sig} ({\bf v}_a - {\bf v}_b ) \cdot {\bf j} }{\bar{\rho}_{ab} },
\label{eq:Piab}
\end{equation}
where $K \sim 0.5$ is a constant, $v_{sig}$ is a signal velocity,  and ${\bf
j}$ is a unit vector from particle $b$ to particle $a$.  The density
$\bar{\rho}_{ab} = \frac12 (\rho_a + \rho_b)$ is an average density.
\begin{equation}
{\bf j } = \frac{ {\bf r}_{ab} }{ |{\bf r}_{ab}| }.
\end{equation}
In the relativistic case it was necessary to replace the velocity by the
momentum and in the relativistic momentum use the velocity along the line of
sight of the two particles in order to guarantee that the viscous dissipation
makes a positive contribution to the thermal energy and therefore to the
entropy. We will see that similar ideas are required here.

The dissipation in the energy equation can be taken as 
\begin{equation}
\Omega_{ab} = - \frac{ K v_{sig} ( e^*_a - e^*_b) {\bf j} }{\bar{\rho_{ab} } },
\label{eq:omegaab}
\end{equation}
where $e^*$ is an energy quantity which is related to $\widehat \epsilon$. Its precise form will now be deduced by considering the rate of change of thermal energy.

 From (\ref{eq:ufrome}) using the SPH equations for the rate of change of  velocity, energy, magnetic field and density we find that the magnetic terms cancel leaving
\begin{eqnarray}
\frac{du_a}{dt} & = & \frac{P_a}{\rho_a^2} \sum_b m_b v_{ab}^i \frac{\partial
W_{ab}}{\partial x_a^i} \nonumber \\
& + & \sum_bm_b  \frac{K v_{sig}}{ \bar{\rho_{ab} } }  \left[ (e^*_a - e^*_b)
\right. \nonumber \\
& & \hspace{2cm} \left. - ({\bf v}_a \cdot {\bf j} )({\bf v}_a - {\bf v}_b ) \cdot
{\bf j} \right) ] |{\bf r}_{ab} |F_{ab}, \nonumber \\
\label{eq:uthermterms}
\end{eqnarray}
where $F_{ab} {\bf r}_{ab} = \nabla W_{ab} $, and $F_{ab} \le 0$ is an even function of $r_{ab}$.

The first term on the right hand side is the adiabatic change in thermal energy
due to the expansion or compression of the gas.  This term does not change the
entropy. The second term is the contribution to the change in thermal energy
due to viscous dissipation, thermal conduction and ohmic heating.  Only the
first and last of these must contribute a non negative quantity to the change
in the thermal energy.  Heat conduction can either increase or decrease the
thermal energy of an element of fluid.  However, all three must contribute a
positive quantity to the change of entropy of the system. The proof that the
contribution to the entropy is positive is given in appendix \ref{sec:entropy}.
The terms due to
viscous dissipation and ohmic dissipation must be negative definite because
$F_{ab} \le 0$.

It is natural to try and construct $e^*$ using the terms in $\widehat \epsilon$,
namely 
\begin{equation}
\widehat \epsilon = \frac12 v^2 + u + \frac{B^2}{2 \mu_0 \rho}.
\label{eq:ehat2}
\end{equation}

\subsection{Viscous dissipation}

The kinetic energy combined with the velocity terms in (\ref{eq:ehat2}) is not negative
definite  and in this form cannot be the correct viscous dissipation.  We get a
positive definite viscous dissipation in the velocity terms by choosing
\begin{equation}
e^*_a - e^*_b = \frac12 ( {\bf v}_{a} \cdot {\bf j} )^2 - \frac12 ({\bf v}_{b} \cdot {\bf j} )^2 + u_a - u_b + \frac{B_a^2}{2 \mu_0 \rho_a} - \frac{B_b^2}{2 \mu_0 \rho_b}.
\end{equation}
We can then write the velocity terms in (\ref{eq:uthermterms}) as
\begin{equation}
- \frac12  \left [ ({\bf v}_a \cdot {\bf j} ) -  ({\bf v}_b \cdot {\bf j}) \right ]^2,
\label{eq:vdiss}
\end{equation}
which is negative definite.  Recalling that $F_{ab} \le 0$ the viscous
contribution to the thermal energy is therefore positive.  Note that the
combination $(e^*_a - e^*_b)$ is not a simple difference because both $e^*_a$ and
$e^*_b$ involve ${\bf j }$ which depends on both particles.

\subsection{Ohmic dissipation}

The magnetic term is wrong because it can be positive or negative.  In
addition, it depends on the total field, whereas in a shock we would expect it
to involve only the component perpendicular to the shock. We therefore replace
the magnetic energy term by using the component of the field perpendicular to
the line of sight of the two particles $a$ and $b$,  then we have

\begin{eqnarray}
e^*_a - e^*_b & = & \frac12 ({\bf v}_{a} \cdot {\bf j} )^2 - \frac12( {\bf v}_{b}
\cdot {\bf j} )^2 + u_a - u_b \nonumber \\
& + & \frac{1}{2 \mu_0 \bar{\rho}_{ab} } \left [ B^2_a -
({\bf B}_a \cdot {\bf j} )^2 -  B^2_b + ({\bf B}_b \cdot {\bf j} )^2 \right  ].
\end{eqnarray}

The magnetic term is still not negative definite.  To make it negative definite we need to add a term 
\begin{equation}
\frac{1}{\mu_0 \bar{\rho}_{ab}} \left\{  ({\bf B}_a \cdot {\bf j})\left [ ({\bf
B}_a \cdot {\bf j} )  -( {\bf B_b \cdot j }) \right ] - {\bf B}_a \cdot [ {\bf
B}_a - {\bf B}_b ] \right\}.
\end{equation} 
 With this new term added the magnetic contribution to the thermal energy  becomes
\begin{equation}
-\frac{1}{2\bar{\rho}_{ab} \mu_0 } \left[{\bf B}_{ab}^2 - ({\bf B}_{ab} \cdot {\bf
j} )^2 \right],
\label{eq:Bdiss}
\end{equation}
which is negative definite and, when combined with $F_{ab}$, gives a positive contribution to the thermal energy change.  

The interpretation of the extra magnetic terms is quite simple : when currents
are present, and the conductivity is finite, the induction equation requires an
extra term as in (\ref{eq:induction}). The contribution to the rate of change
of thermal energy from the new term is
\begin{equation}
-\frac{{\bf B}_a}{\mu_0} \cdot \sum_b m_b \frac{K v_{sig}}{\bar{\rho_{ab}
}^2}\left  [ {\bf B}_{ab}  - {\bf j} ({\bf B}_{ab} \cdot  {\bf j} ) \right ] r_{ab} F_{ab}.
\label{eq:bdotterm}
\end{equation}

The term which must be added to the induction equation for consistency can be
deduced by noting that the expression for the rate of change of thermal energy
(\ref{eq:ufrome}) has a magnetic term

\begin{equation} 
 - \frac{d}{dt} \left ( \frac{ B^2}{2 \rho \mu_0 } \right ) = - \frac{ {\bf B} }{\rho \mu_0} \frac{ d {\bf B}}{dt} + \frac{B^2}{\rho^2 \mu_0} \frac{d \rho}{dt}.
\end{equation}
Comparing the first term with (\ref{eq:bdotterm}) we find that the SPH induction equation requires a term
\begin{equation}
\left.\frac{d\bB_a}{dt}\right\vert_{diss} = \rho_a \sum_b m_b \frac{K v_{sig}}{\bar{\rho}_{ab}^2}({\bf j} \times ( {\bf B}_{ab} \times {\bf j})r_{ab} F_{ab},
\label{eq:jcrossbab}
\end{equation}
and we expect that the continuum version of this term should be some approximation to 
\begin{equation}
-\nabla \times (\eta \nabla \times {\bf B} ),
\end{equation}
which when $\eta$ is constant is
\begin{equation}
\eta\left[\nabla^2 {\bf B }  -  \nabla ( \nabla \cdot {\bf B} ) \right].
\end{equation}

By replacing the summation in (\ref{eq:jcrossbab}) by an integral, and expanding
in a Taylor series about ${\bf r}_a$, and assuming that $v_{sig}$ is constant, 
we find that (\ref{eq:jcrossbab})  is proportional to 
 \begin{equation}
 Kv_{sig} h \left  [  \nabla^2 {\bf B} - \frac23 \nabla (\nabla \cdot {\bf B} ) \right ] ,
\end{equation}
which is similar to the exact equation with ohmic diffusivity $\eta \propto K v_{sig} h$.

\subsection {The signal velocity}
\label{sec:vsig}

We refer the reader to a general discussion of signal velocities in
\citet{monaghan97} and \citet{cm97}.  The key point is that it is the relative speed of signals from moving observers at the positions of particles $a$ and $b$ when the signals are sent along the line of sight.   If there are no magnetic fields a good estimate of this signal velocity is
\begin{equation}
v_{sig} = c_a + c_b -  \beta {\bf v}_{ab} \cdot {\bf j},
\label{eq:vsig}
\end{equation}
where $c_a$ denotes the speed of sound of particle $a$ and $\beta \sim 1$. The signal
velocity  is larger when the particles are approaching each other and in practice,
the effects of shocks can be included by choosing  $\beta = 2$ (however when the
artificial dissipation switch discussed in \S\ref{sec:avswitch} is used we find it is
better to set $\beta = 1$ due to the stronger source term).  If there are
magnetic fields then a variety of other waves are possible.  The fastest wave in a
static medium along the x axis has speed
\begin{equation}
\frac12 \left [    \sqrt{ c^2 + \frac{B^2}{\rho \mu_0 } + \frac{2 B^x c}{ \sqrt{\rho
\mu_0} } } + \sqrt{ c^2 + \frac{B^2}{\rho \mu_0 } -\frac{2 B^x c}{ \sqrt{\rho
\mu_0} } }   \right ].
\end{equation}
A natural generalization of (\ref{eq:vsig}) for the case of magnetic fields  is to take
\begin{equation}
v_{sig} = v_a + v_b - \beta  {\bf v}_{ab} \cdot {\bf j},
\label{eq:vsigmag}
\end{equation}
where
\begin{equation}
v_a = \frac12 \left [    \sqrt{ c_a^2 + \frac{B_a^2}{\rho_a \mu_0 } + \frac{2
{\bf B}_a \cdot {\bf j} c_a}{ \sqrt{\rho_a \mu_0} } } + \sqrt{ c_a^2 +
\frac{B_a^2}{\rho_a \mu_0 } -\frac{2 {\bf B}_a \cdot {\bf j}  c_a}{ \sqrt{\rho_a
\mu_0} } }   \right ],
\end{equation}
with a similar equation for $v_b$.

\subsection {Artificial dissipation switch}
\label{sec:avswitch}
 A switch to reduce the artificial viscosity away from shocks is given by
\citet{mm97}. Using this switch together with the suggestions of
\citet{balsara95} in multi-dimensional simulations can virtually eliminate the
problematic effects of using an artificial dissipation in SPH. 
 
 The key idea is to regard the dissipation parameter $K$ (c.f. equation
\ref{eq:Piab}) as a particle property. This can then be evolved along with the
fluid equations according to
\begin{equation}
\frac{dK_a}{dt} = -\frac{K_a-K_{min}}{\tau_a} + \mathcal{S}_a,
\end{equation}
such that in the absence of sources $\mathcal{S}$, $K$ decays to a value
$K_{min}$ over a timescale $\tau$. The timescale $\tau$ is calculated
according to
\begin{equation}
\tau = \frac{h}{\mathcal{C}v_{sig}},
\end{equation}
where $h$ is the particle's smoothing length, $v_{sig}$ is the maximum signal
propagation speed at the
particle location and $\mathcal{C}$ is a dimensionless parameter with value $0.1
< \mathcal{C} <
0.2$. We conservatively use $\mathcal{C}=0.1$ which means that the value of $K$ decays to $K_{min}$ over
$\sim 5$ smoothing lengths.

 The source term $\mathcal{S}$ is chosen such that the artificial dissipation grows
as the particle approaches a shock front. We use
\begin{equation}
\mathcal{S} = \mathrm{max}(-\nabla\cdot\bv, 0),
\end{equation} 
such that the dissipation grows in regions of strong compression. Following \citet{mm97} where
the ratio of specific heats $\gamma$ differs from 5/3 (but not for the isothermal case), we multiply $\mathcal{S}$ by a factor
\begin{equation}
\left[\mathrm{ln}\left( \frac{5/3 +1}{5/3-1} \right)\right] /
\left[\mathrm{ln}\left( \frac{\gamma+1}{\gamma-1}\right)\right]
\end{equation}
 
 Note that our source term is a factor of two times larger than the term used by
\citet{mm97} since our dissipation parameter $K$ is required to be of order $K \sim 0.5$ at a shock front,
whilst the usual SPH artificial viscosity parameter $\alpha$ is of order unity.
 We prefer this stronger source term since it provides sufficient damping in the
\citet{sod78} hydrodynamic shock tube problem and in
the MHD shock tube tests we describe in this paper (ie. $K_{max} \sim 0.5$ for
these problems).

In order to conserve momentum the average value $\bar{K}=0.5(K_a + K_b)$
is used in equation (\ref{eq:Piab}) or (\ref{eq:omegaab}). A lower limit of
$K_{min}=0.05$ is used to preserve order away from shocks (note that this is an
order of magnitude reduction from the usual value of $K=0.5$ everywhere).

 The numerical tests in \S\ref{sec:1Dtests} demonstrate that use of this limiter gives a significant
reduction in dissipation away from shocks whilst preserving the
shock-capturing ability of the code.

\section{Instability correction}
\label{sec:instability}
 The tensile instability is corrected via the method proposed by
\citet{monaghan00}. The idea is add a small term which prevents
particles clumping under negative stress. The momentum equation (\ref{eq:tensor}) becomes
\begin{eqnarray}
\frac{dv^i_a}{dt} & = & \smb\left\{\left(\frac{S_{ij}}{\rho^2}\right)_a +
\left(\frac{S_{ij}}{\rho^2}\right)_b \right.\nonumber \\
& + & \left. R\left[ \left(\frac{B_i B_j}{\rho^2}\right)_a + \left(\frac{B_i
B_j}{\rho^2}\right)_b \right]\right\}\pder{W_{ab}}{x_{j,a}},
\end{eqnarray}
where $R$ is a function which increases as the particle separation decreases,
given by
\begin{equation}
\label{eq:R}
R = -\frac{\epsilon}{2\mu_0}\left(\frac{W_{ab}}{W(\Delta p)}\right)^n,
\end{equation}
where $W$ is the SPH kernel and W($\Delta p$) is the kernel evaluated at the average particle spacing.
Further details of the derivation of this term are given in appendix
\ref{sec:appendix1}.
For all the simulations
presented here the particles are setup with $h = 1.5 \Delta p$ and
therefore in (\ref{eq:R}) we compute the kernel in the denominator using $\Delta
p/h = 1/1.5$. We use $\epsilon = 0.8$ and $n=5$ throughout this paper and in one
dimension we apply the correction only in the
x-direction. 

 Note that where the
total energy equation (\ref{eq:sphenergy}) is used, the source term
\begin{equation}
\left.\frac{d\widehat{\epsilon}_a}{dt}\right\vert_{src} = \smb v_a^i R \left[ \left(\frac{B_i B_j}{\rho^2}\right)_a + \left(\frac{B_i
B_j}{\rho^2}\right)_b \right]\pder{W_{ab}}{x_{j,a}},
\end{equation}
is added for consistency. 

We show in \S\ref{sec:1Dtests} that this correction term very effectively removes the
tensile instability with few side effects.


\section{Timestepping}
\label{sec:timestep}
We integrate the SPMHD equations using a simple midpoint
predictor-corrector method. Quantities are
predicted according to
\begin{equation}
\bA^{1/2} = \bA^0 + \frac{\Delta t}{2} \left(\frac{d\bA}{dt}\right)^{-1/2},
\end{equation}
where
\begin{equation}
\bA = [x, v_x, v_y, v_z, \rho, \hat{\epsilon}, B_y, B_z, h, K]^T,
\end{equation}
with the energy $\hat{\epsilon}$ interchangeable for the thermal energy $u$. Note that we evolve the
smoothing length alongside the particle equations as discussed in
\S\ref{sec:sphforms} and
that the dissipation parameter $K$ is also evolved in accordance with the
switch discussed in \S\ref{sec:avswitch}. The density is also included since in
this paper the continuity equation (\ref{eq:sphcty}) is integrated rather than using the density summation
(\ref{eq:rhosum}).

The rates of change $d\bA/dt$ of these quantities are then computed via the SPH
summations using the predicted values $\bA^{1/2}$. The corrector step is given by
\begin{equation}
\bA^* = \bA^0+ \frac{\Delta t}{2} \left(\frac{d\bA}{dt}\right)^{1/2},
\end{equation}
and
\begin{equation}
\bA^{1} = 2 \bA^* - \bA^0.
\end{equation}

 The timestep is determined by the Courant condition
\begin{equation}
\mathrm{dt_c} = C_{cour}\mathrm{min}\left(\frac{\bar{h}_{ab}}{v_{sig,ab}}\right)
\label{eq:dtc}
\end{equation}
where $\bar{h}_{ab} = 0.5(h_a + h_b)$, the signal velocity $v_{sig}$ is given
by equation (\ref{eq:vsigmag}), except that we use 
\begin{equation}
v_{sig,ab} = v_a + v_b + \beta\vert \bv_{ab}\cdot \mathbf{j} \vert
\end{equation}
with $\beta = 1$ when $\bv_{ab}\cdot \mathbf{j} > 0$ (ie. where the dissipation
terms are not applied). The minimum in (\ref{eq:dtc}) is taken over all particle
interactions and in this paper we use $C_{cour} = 0.8$.

Although this condition is sufficient for all of the simulations described here, 
in general it is necessary to pose the additional constraint from the forces
\begin{equation}
\mathrm{dt_f} = \mathrm{min}\left(\frac{h_a}{\vert \mathbf{a}_a \vert}\right)^{1/2},
\end{equation}
where $\mathbf{a}_a$ is the acceleration on particle $a$.


\section{Numerical tests in one dimension}
\label{sec:1Dtests}
 The numerical scheme described in this paper has been tested on a variety of one
dimensional problems. In order to demonstrate that SPMHD gives good results on
problems involving discontinuities in the physical variables we present results of
standard problems used to test grid-base MHD codes (e.g. \citealt{sea92,dw94,rj95,balsara98,dw98}).
The advantages of SPMHD are the simplicity with which these results can be obtained and the
complete absence of any numerical grid. Further tests (MHD waves) are given in
paper II since they incorporate the use of the variable smoothing
length terms.

\subsection{Implementation}
The particles are allowed to move in one dimension
only, whilst the velocity and magnetic field are allowed to vary in three
dimensions. We use equal mass particles such that density changes correspond
to changes in particle spacing. Unless otherwise indicated in this paper we integrate the continuity
equation (\ref{eq:sphcty}), the momentum equation (\ref{eq:tensor}), the
total energy equation (\ref{eq:sphenergy}) and the induction equation
(\ref{eq:sphinduction}). This is
the most efficient implementation of the SPMHD equations since it does not
require an extra pass over the particles to calculate the density
via the summation (\ref{eq:rhosum}).
Similar results to those shown here are also obtained when the thermal energy
equation is integrated instead of the total energy. Additionally we note that whilst evolving the flux per unit
mass (\ref{eq:Bevolsph2}) instead of the flux density (\ref{eq:sphinduction}) does 
not exactly maintain $\divB = 0$ in one dimension, the associated errors are small and
hence we also find in this case that the results are similar. 
 Unless otherwise indicated the tests presented here are all performed with the
artificial dissipation switch discussed in \S\ref{sec:avswitch} turned on with minimum
dissipation parameter $K_{min} = 0.05$. This results in very little dissipation
away from shock fronts.

\subsubsection{Scaling}
  The magnetic field variable is scaled in units such
that the constant $\mu_0$ is unity and numerical quantities are
dimensionless. Note that the magnetic
flux density $\bB$ has dimensions
\begin{equation}
[\bB] = \frac{[mass]}{[time][charge]},
\end{equation}
whilst $\mu_0$ has dimensions
\begin{equation}
[\mu_0] = \frac{[mass][length]}{[charge]^2}.
\end{equation}
Choosing mass, length and time scales of unity and specifying $\mu_0 = 1$ therefore
defines the unit of charge. Re-scaling of the magnetic field variable to physical
units requires multiplication of the code value by a constant
\begin{equation}
\bB_{physical} = \left\{\frac{\mu_0 [mass]}{[length][time]^2}\right\}^{1/2}
\bB_{numerical}.
\end{equation}
For example, in cgs units, with mass, length and time scales of unity the magnetic flux density in Gauss is given by
\begin{equation}
\bB_{cgs} = (4\pi)^{1/2} \bB_{numerical}.
\end{equation}

\subsubsection{Initial conditions}
\label{sec:initconds}
 Integration of the continuity equation (\ref{eq:sphcty}) requires some smoothing of the
initial conditions and we follow \citet{monaghan97} such that when an initial
quantity $A$ is discontinuous it is smoothed according to the rule
\begin{equation}
A = \frac{A_L + A_R e^{x/d}}{1 + e^{x/d}}
\end{equation}
where $A_L$ and $A_R$ are the uniform left and right states with respect to the
origin and $d$ is taken as half of the largest initial particle separation at
the interface (ie. the particle separation on the low density side). Note
however that we \emph{do not} smooth the initial velocity profiles except in the
rarefaction test. Where the
initial density is smoothed the particles are spaced according to the rule
\begin{equation}
\rho_a(x_{a+1} - x_{a-1}) = 2 \rho_R \Delta_R
\end{equation}
where $\Delta_R$ is the particle spacing to the far right of the origin with
density $\rho_R$. Note that initial smoothing lengths are set according to the
rule $h \propto 1/\rho$ and are therefore also smoothed. Where the total energy
$\hat{\epsilon}$ is integrated we smooth the basic variables $u$ and
$\bB$ and construct the total energy from (\ref{eq:ehat}). There is some
inconsistency in the smoothing in this case as it is not possible to self consistently construct smooth
profiles of $\rho$, $u$, $B_y$, $B_z$ and $\hat{\epsilon}$ with the above smoothing. This can cause a small
glitch in the initial conditions when the total energy equation is used.

 Such smoothing of the initial conditions can be avoided altogether if the density summation (\ref{eq:rhosum}) is used,
particularly if the smoothing length is updated self-consistently with the
density. This is demonstrated in paper II.

\subsubsection{Boundaries}
Boundary conditions are implemented in one dimension by simply fixing the
properties of the 6 particles closest to each boundary. Where the initial
velocities of these particles are non-zero their positions are evolved
accordingly and a particle is removed from
the domain once it has crossed the boundary. Where the distance between the
closest particle and the boundary is more than the initial particle spacing
a new particle is introduced to the domain. Hence for inflow or outflow boundary
conditions the resolution changes throughout the simulation.

\subsubsection{Dissipative terms}
 Direct application of the dissipation terms described in \S\ref{sec:dissipation} provides no smoothing
for the velocity in the $y$ (and $z$) directions since we use $({\bf v}_a - {\bf v}_b ) \cdot {\bf j}$ in
equation (\ref{eq:Piab}) and the particles are restricted to move along the $x$
axis only. If the particles were allowed to move in the $y$($z$) direction (with velocity
$v_y$($v_z$) ) such smoothing would naturally be present. Therefore in the simulations
presented in this paper we use
\begin{equation}
\Pi_{ab} = - \frac { K v_{sig} ({\bf v}_a - {\bf v}_b ) \cdot {\bf \hat{v}} }{\bar{\rho}_{ab} },
\end{equation}
where $\bf\hat{v}$ is a unit vector along the direction of the particle velocities given by
\begin{equation}
\hat{\bv} = \frac{\bv_a - \bv_b}{\vert \bv_a - \bv_b\vert}.
\end{equation}
The term in the momentum equation is then
\begin{equation}
\smb \Pi_{ab} \hat{\bv} G_{ab},
\end{equation}
where we note that the gradient of the kernel may be written as $\gwab =
\mathbf{j} G_{ab}$, which we replace with $\hat{\bv} G_{ab}$ where $G_{ab}$  is a
scalar function (note that $G_{ab} = F_{ab} h/\vert \br_{ab}\vert$).

In the dissipative energy we use
\begin{equation}
e^*_a = \frac12 (v_a\cdot\hat{\bv})^2 + u_a + \frac12 (\bB_a\cdot {\bf j})^2,
\end{equation}
with the contribution to the thermal energy equation from the kinetic terms given by
\begin{equation}
-\frac12\left[(\bv_a\cdot\hat{\bv}) - (\bv_b\cdot\hat{\bv})\right]^2.
\end{equation}
Note that the dissipative terms are only applied where $\bv_{ab}\cdot{\bf j} > 0$.

\subsection{Simple advection test}
 This simple test is described in \citet{eh88} and in \citet{sea92} and measures
the ability of an algorithm to advect contact discontinuities. A square pulse
of transverse magnetic field is setup and advected a distance of five times its
width with the pressure terms switched off. The current density $\mathbf{J}$ is calculated in order to
ascertain that the method does not produce sign reversals or anomalous extrema
in this quantity. In SPH we compute this quantity using
\begin{equation}
\mathbf{J}_a = \curl \bB_a = \smb (\bB_a-\bB_b) \times \gwab.
\end{equation}

 We perform this test simply by using a magnetic pressure that
is negligible compared to the gas pressure. We setup 100 particles placed evenly
along the x axis with constant velocity in the positive x-direction and use a pulse
that is initially 50 particle spacings wide. The pulse is not initially smoothed
in any way and periodic boundary conditions are enforced using ghost particles
(this is also a good test of the periodic boundary conditions since the
particles are continually crossing the domain). 

\begin{figure*}[t]
\begin{center}
\begin{turn}{270}\epsfig{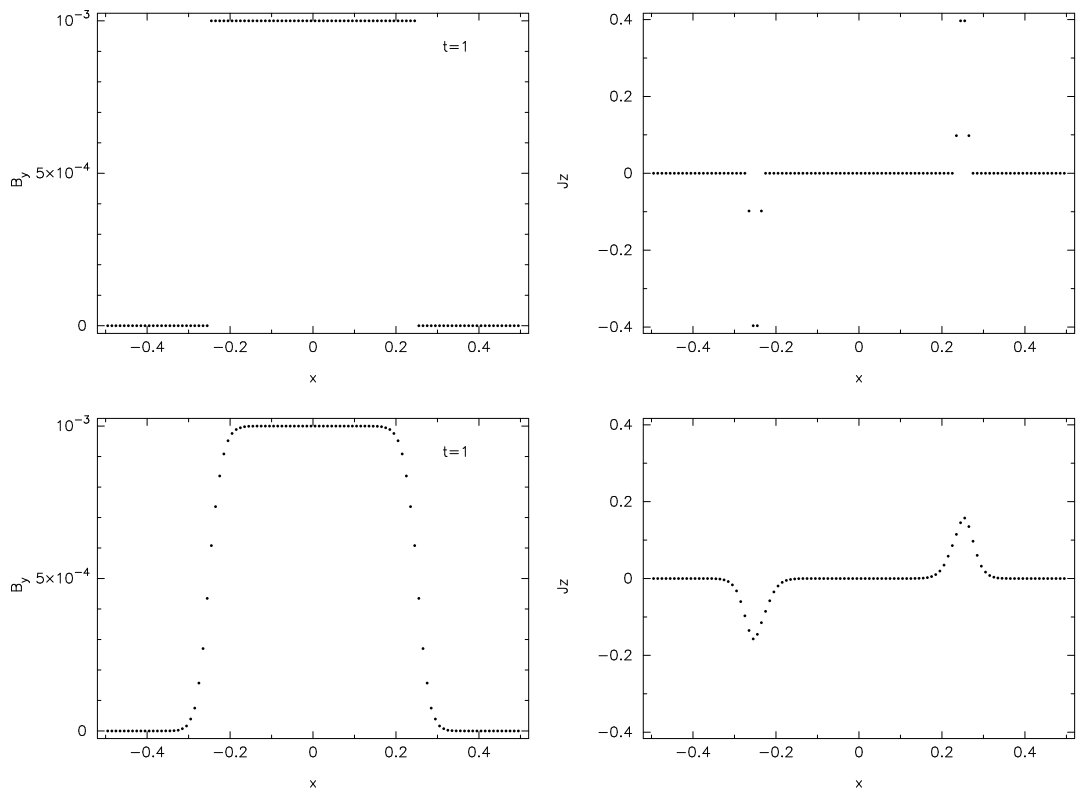}\end{turn}
\caption{Results of the advection of a square pulse of transverse magnetic field
50 particle separations wide a distance of five times its width. In the absence
of dissipative terms the discontinuities are kept to less than a
particle spacing (top) and the current density (top right) shows almost no spread
(analytically this is a delta function at each discontinuity). With the
dissipative terms included with the \citet{mm97} switch a small amount of smoothing
is observed (bottom panels)}
\label{fig:advection}
\end{center}
\end{figure*}

 The SPMHD results are shown in Figure
\ref{fig:advection} after advecting the pulse a distance of five times its width (in this case 5
crossings of the computational domain). The top panel shows the results with the
artificial dissipation terms turned off. The spread in the discontinuities are kept
to less than a particle spacing, showing no visible dispersion or diffusion
whatsoever, suggesting that SPH indeed handles contact discontinuities very
well. The current density, which is analytically given by a delta function at
each discontinuity, is also computed very well by the SPH approximation
(see \citealt{monaghan92}). When the dissipative terms are turned on using the
switch (\S\ref{sec:avswitch}) a small smoothing of the field is observed (bottom
panels), however this
still compares favourably with the schemes shown in \citet{sea92}.

\subsection{Shock tubes}
 The first shock tube test we perform was first described by \citet{bw88} and is the MHD analog of the
\citet{sod78} shock tube problem. The problem consists of a
discontinuity in pressure, density, transverse magnetic field and
internal energy initially located at the origin. As time develops complex shock structures
develop which only occur in MHD because of the different wave types.
Specifically the \citet{bw88} problem contains a compound wave consisting of a
slow shock attached to a rarefaction wave. The existence of such intermediate
shocks was contrary to the expectations of earlier theoretical studies
\citep{bw88}. This problem is now a standard test for any astrophysical
MHD code and has been used by many authors (e.g. \citealt{sea92,dw94,rj95,balsara98})
                        
\begin{figure*}[t]
\begin{center}
\epsfig{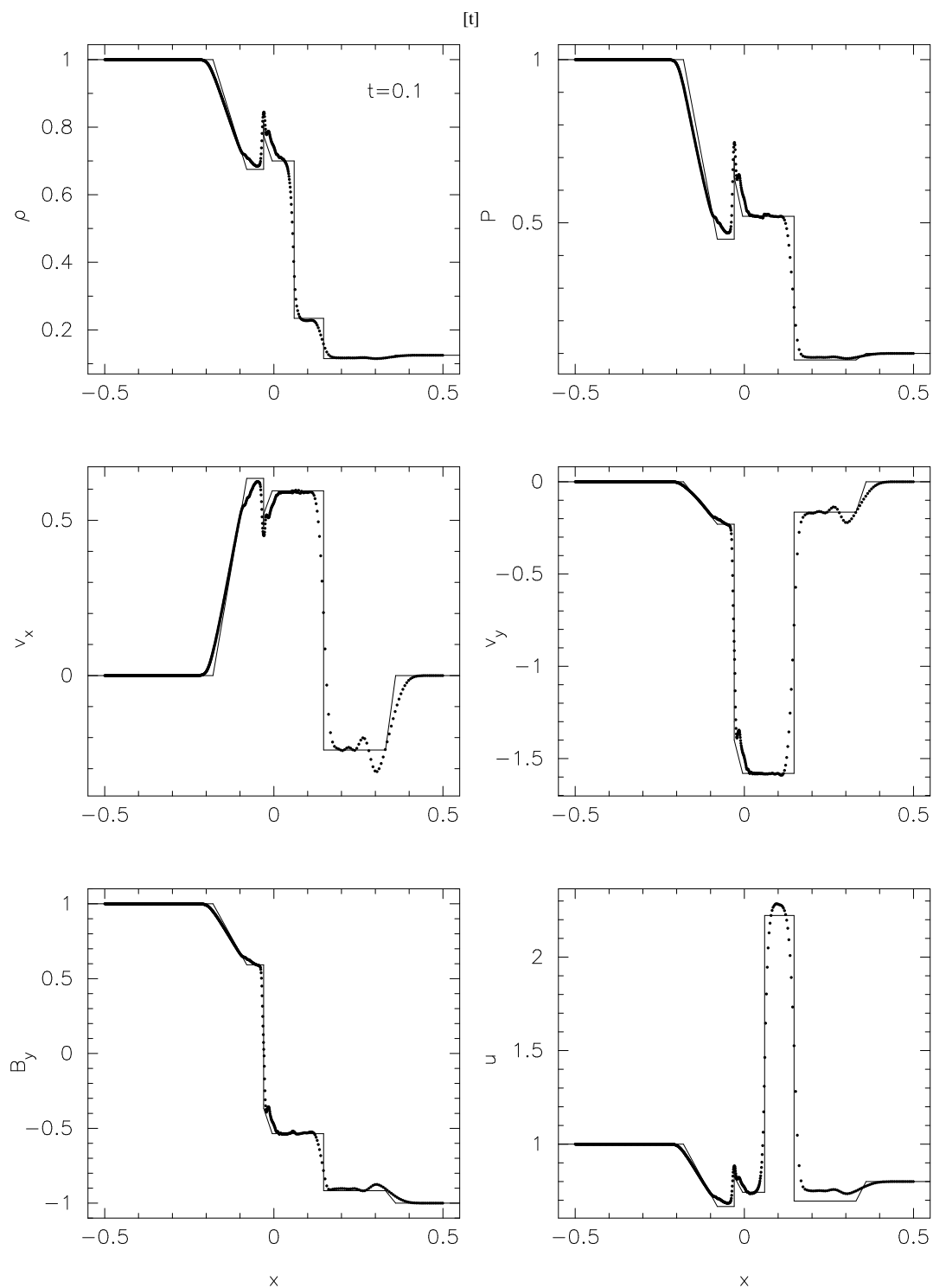}
\caption{Results of the \citet{bw88} shock tube test. To the left of the origin
the initial state is $(\rho,P,v_x,v_y,B_y) = [1,1,0,0,1]$ whilst to the right
the initial state is
$(\rho,P,v_x,v_y,B_y)=[0.125,0.1,0,0,-1]$ with
$B_x = 0.75$ everywhere and $\gamma=2.0$. Profiles of density, pressure, $v_x$, $v_y$,
thermal energy and $B_y$ are shown at time $t=0.1$. Points indicate
the SPMHD particles whilst the numerical solution from \citet{balsara98} is given by the solid lines. The artificial dissipation switch with $K_{min}=0.05$ is used.}
\label{fig:briowu}
\end{center}
\end{figure*}

 We set up the problem using approximately 800 equal mass particles in the domain $x =
[-0.5, 0.5]$. Initial conditions to the left of the discontinuity (hereafter the left
state) are given by $(\rho,P,v_x,v_y,B_y) = [1,1,0,0,1]$ and
conditions to the right (hereafter the right state) are given by $(\rho,P,v_x,v_y,B_y)=[0.125,0.1,0,0,-1]$ with
$B_x = 0.75$ and $\gamma=2.0$. The results are shown in Figure \ref{fig:briowu}
at time $t=0.1$ and compare well with the numerical solution in
\citet{balsara98} (solid lines).
Similar results to Figure \ref{fig:briowu} are obtained when the thermal energy
equation is integrated.

\begin{figure*}[t]
\begin{center}
\epsfig{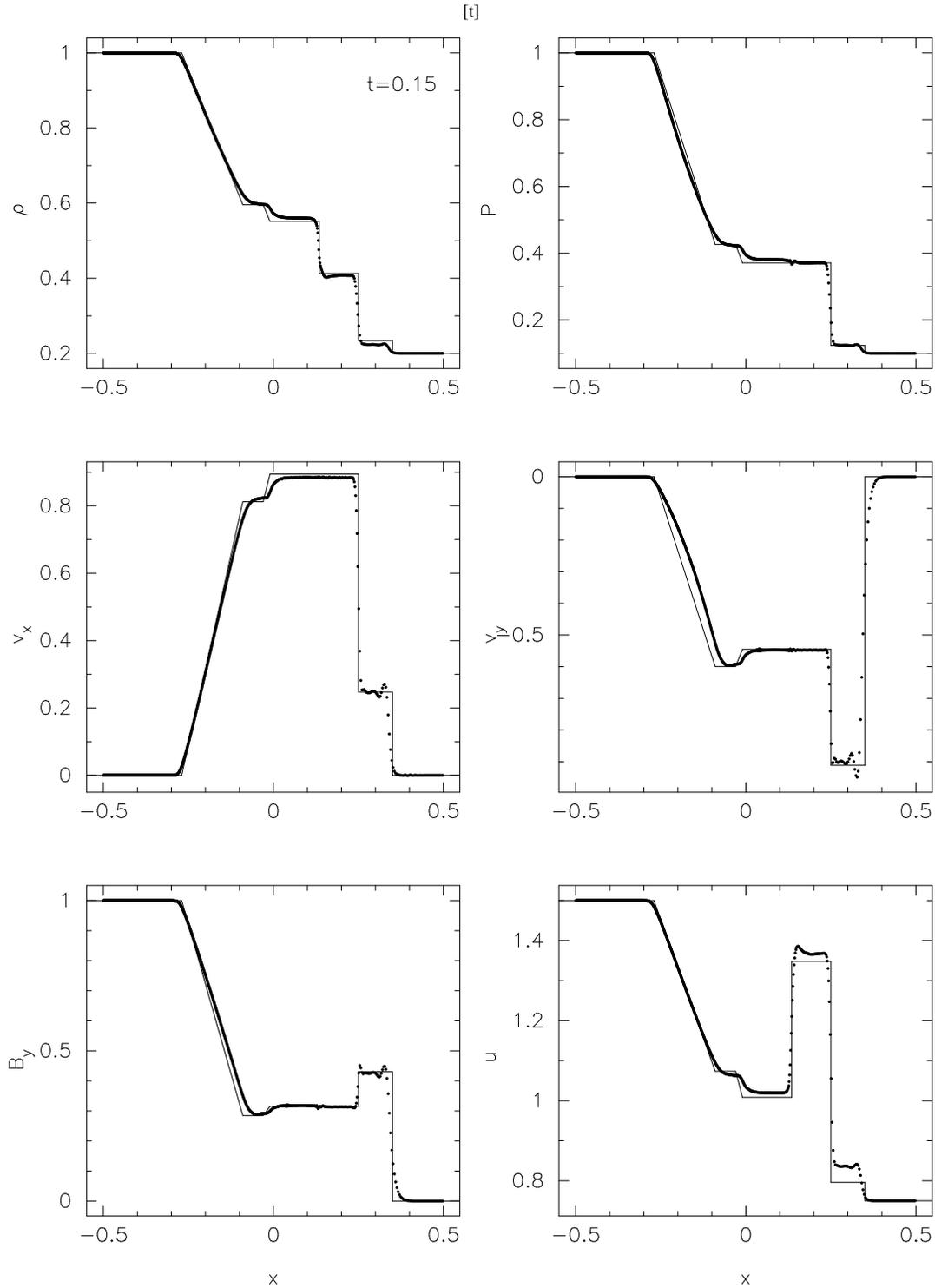}
\caption{Results of the MHD shock tube test with left state
$(\rho,P,v_x,v_y,B_y) = [1,1,0,0,1]$ and the right state
$(\rho,P,v_x,v_y,B_y)=[0.2,0.1,0,0,0]$ with
$B_x = 1$ and $\gamma=5/3$ at time $t=0.15$. The problem illustrates the formation
of a switch-on fast shock and the solution contains both a fast and slow shock.
Solid points indicate the SPMHD particles whilst the exact solution is given by
the solid line. The artificial dissipation switch is
used. Without this switch the fast shock is significantly damped.}
\label{fig:balsara1}
\end{center}
\end{figure*}

 In the second shock tube test (Figure \ref{fig:balsara1}), we demonstrate the usefulness of the
artificial dissipation switch by considering a problem which involves both a fast
and slow shock. We consider the Riemann problem with left state $(\rho,P,v_x,v_y,B_y) = [1,1,0,0,1]$ and the right state
$(\rho,P,v_x,v_y,B_y)=[0.2,0.1,0,0,0]$ with $B_x = 1$ and $\gamma=5/3$. This test
has been used by \citet{dw94}, \citet{rj95} and \citet{balsara98} and illustrates
the formation of a switch-on fast shock.
 Similarly to the previous test we set up the simulation using approximately 800
particles in the domain $x = [-0.5,0.5]$. The results are shown
in Figure \ref{fig:balsara1} at time $t=0.15$ and compare well with the exact
solution given by \citet{rj95} (solid lines). The advantages of the dissipation switch 
are apparent in this problem since it contains both a fast and slow shock. In a run
with dissipation parameter $K=0.5$ everywhere the fast shock is significantly damped.
In Figure \ref{fig:balsara1} we see that the fast shock is resolved, although some small oscillations are
observed behind the shock front. These oscillations can be removed entirely by
using a slightly higher minimum dissipation parameter ($K_{min}=0.1$). 

\begin{figure*}[t]
\begin{center}
\epsfig{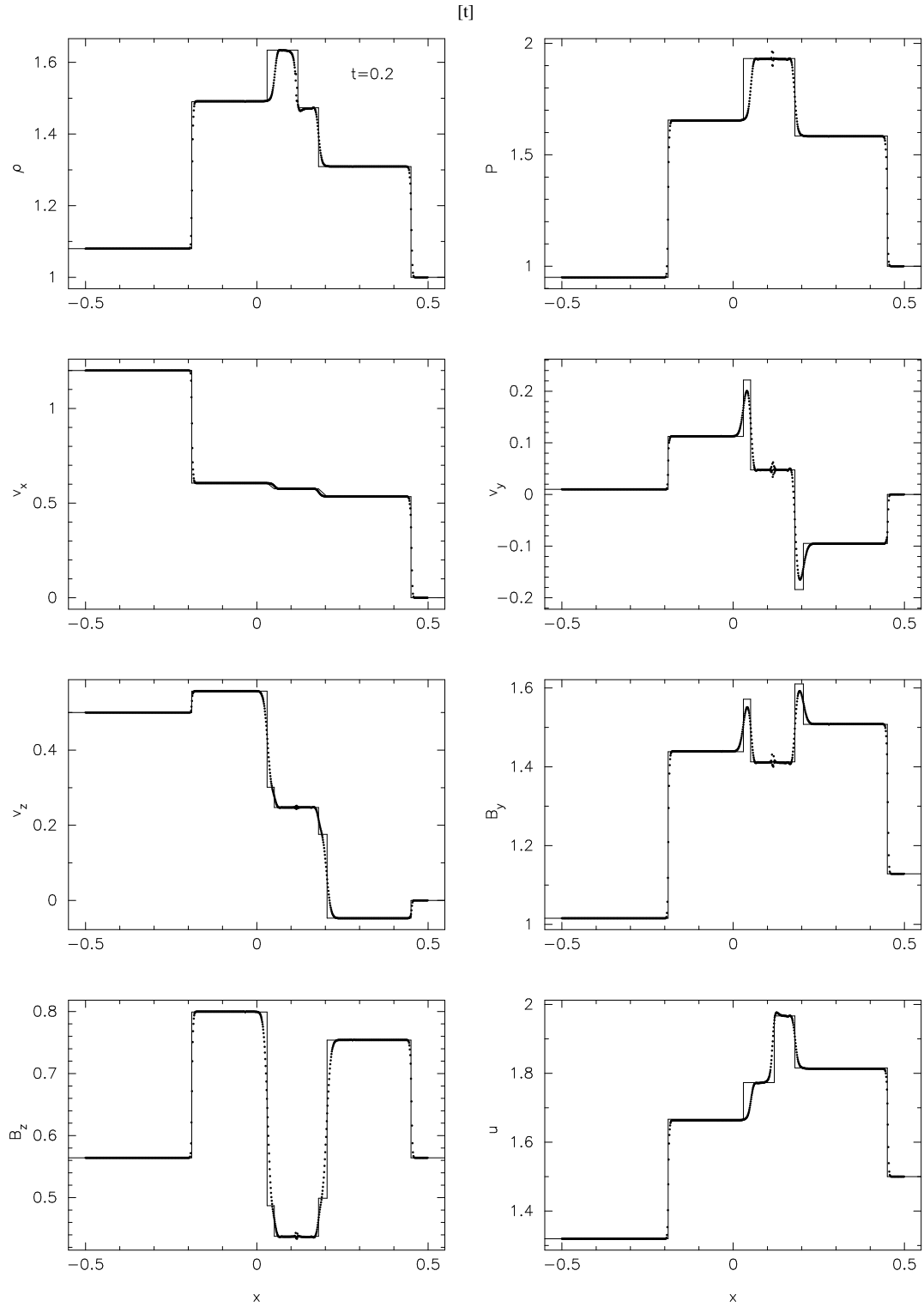}
\caption{Results of the MHD shock tube test with left state
$(\rho,P,v_x,v_y,v_z,B_y,B_z) =
[1.08,0.95,1.2,0.01,0.5,3.6/(4\pi)^{1/2},2/(4\pi)^{1/2}]$ and right state
$(\rho,P,v_x,v_y,v_z,B_y,B_z)=[1,1,0,0,0,4/(4\pi)^{1/2},2/(4\pi)^{1/2}]$ with $B_x =
2/(4\pi)^{1/2}$ and $\gamma=5/3$ at time $t=0.2$. This problem illustrates the formation
of seven discontinuities. The exact solution is given by the solid line whilst
points indicate the positions of the SPMHD particles. The artificial
dissipation switch is used with $K_{min} = 0.05$}
\label{fig:balsara3}
\end{center}
\end{figure*}

 The third test illustrates the formation of seven discontinuities in the same
problem (Figure \ref{fig:balsara3}). The left state is given by $(\rho,P,v_x,v_y,v_z,B_y,B_z) =
[1.08,0.95,1.2,0.01,0.5,3.6/(4\pi)^{1/2},2/(4\pi)^{1/2}]$ and the right state
$(\rho,P,v_x,v_y,v_z,B_y,B_z)=[1,1,0,0,0,4/(4\pi)^{1/2},2/(4\pi)^{1/2}]$ with $B_x =
2/(4\pi)^{1/2}$ and $\gamma=5/3$. Since the velocity in the
x-direction is non-zero at the boundary, we continually inject particles into
the left half of the domain with the appropriate left state properties. The
resolution therefore varies from an initial 700 particles to 875 particles at
$t=0.2$. The results are shown in Figure \ref{fig:balsara3} at time $t=0.2$. The
SPMHD solution
compares extremely well with the exact solution taken from \citet{rj95} (solid
line) and may also be compared with the numerical solution in that paper and in \citet{balsara98}. The thermal energy and density profiles are slightly
improved by our use of the total energy equation. Note that the initial velocity
profiles are not smoothed for this problem, resulting in the small starting error at
the contact discontinuity.

\begin{figure*}[t]
\begin{center}
\epsfig{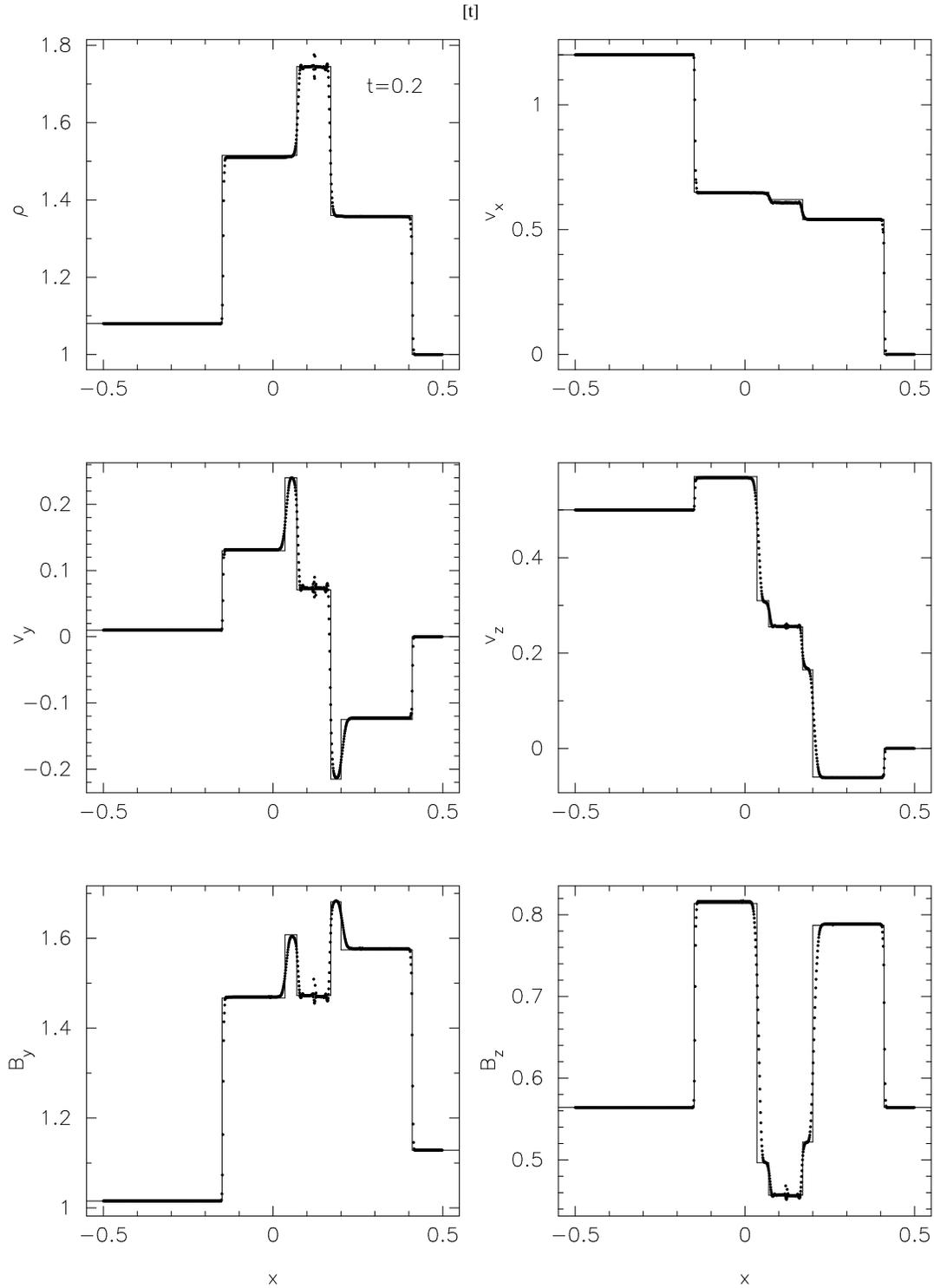}
\caption{Results of the isothermal MHD shock tube test with left state
$(\rho,v_x,v_y,v_z,B_y,B_z) =
[1.08,1.2,0.01,0.5,3.6/(4\pi)^{1/2},2/(4\pi)^{1/2}]$ and right state
$(\rho,P,v_x,v_y,v_z,B_y,B_z)=[1,0,0,0,4/(4\pi)^{1/2},2/(4\pi)^{1/2}]$ with $B_x =
2/(4\pi)^{1/2}$ and an isothermal sound speed of unity at time $t=0.2$. This problem illustrates the formation
of six discontinuities in isothermal MHD. Solid points indicate the position of
the SPMHD particles which may be compared with the exact solution given by the
solid line. The artificial
dissipation switch is used with $K_{min} = 0.05$.}
\label{fig:balsara5}
\end{center}
\end{figure*}

\begin{figure*}[t]
\begin{center}
\epsfig{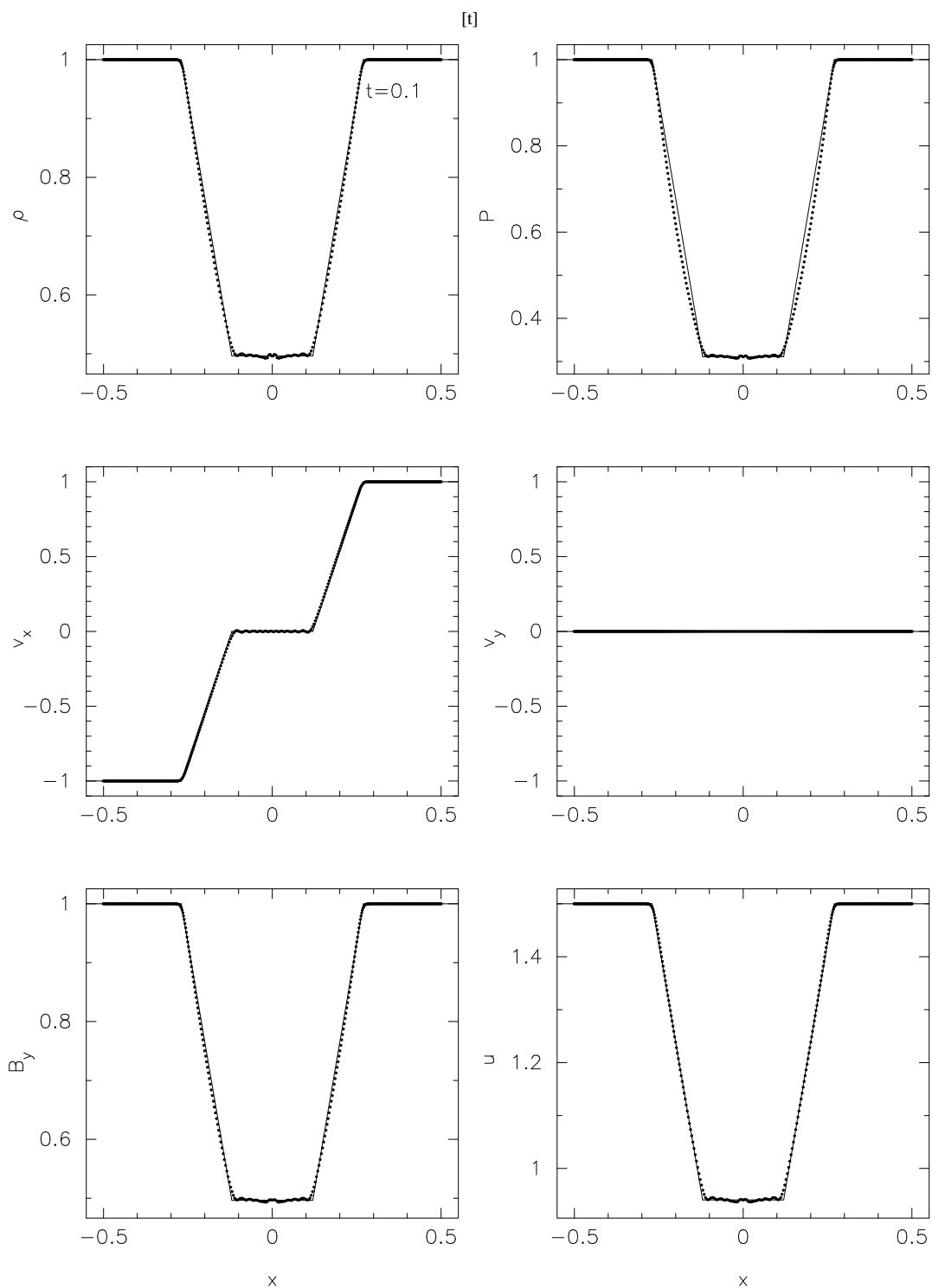}
\caption{Results of the MHD shock tube test with left state
$(\rho,P,v_x,v_y,B_y) =
[1,1,-1,0,1]$ and right state
$(\rho,P,v_x,v_y,B_y)=[1,1,1,0,1]$ with $B_x = 0$ and $\gamma = 5/3$ at time
$t=0.1$. This problem illustrates the formation
of two magnetosonic rarefactions. The exact solution is given by the solid line
whilst points indicate the position of the SPMHD particles. The artificial
dissipation switch is used with $K_{min} = 0.05$.}
\label{fig:rarefaction}
\end{center}
\end{figure*}

\begin{figure*}[t]
\begin{center}
\epsfig{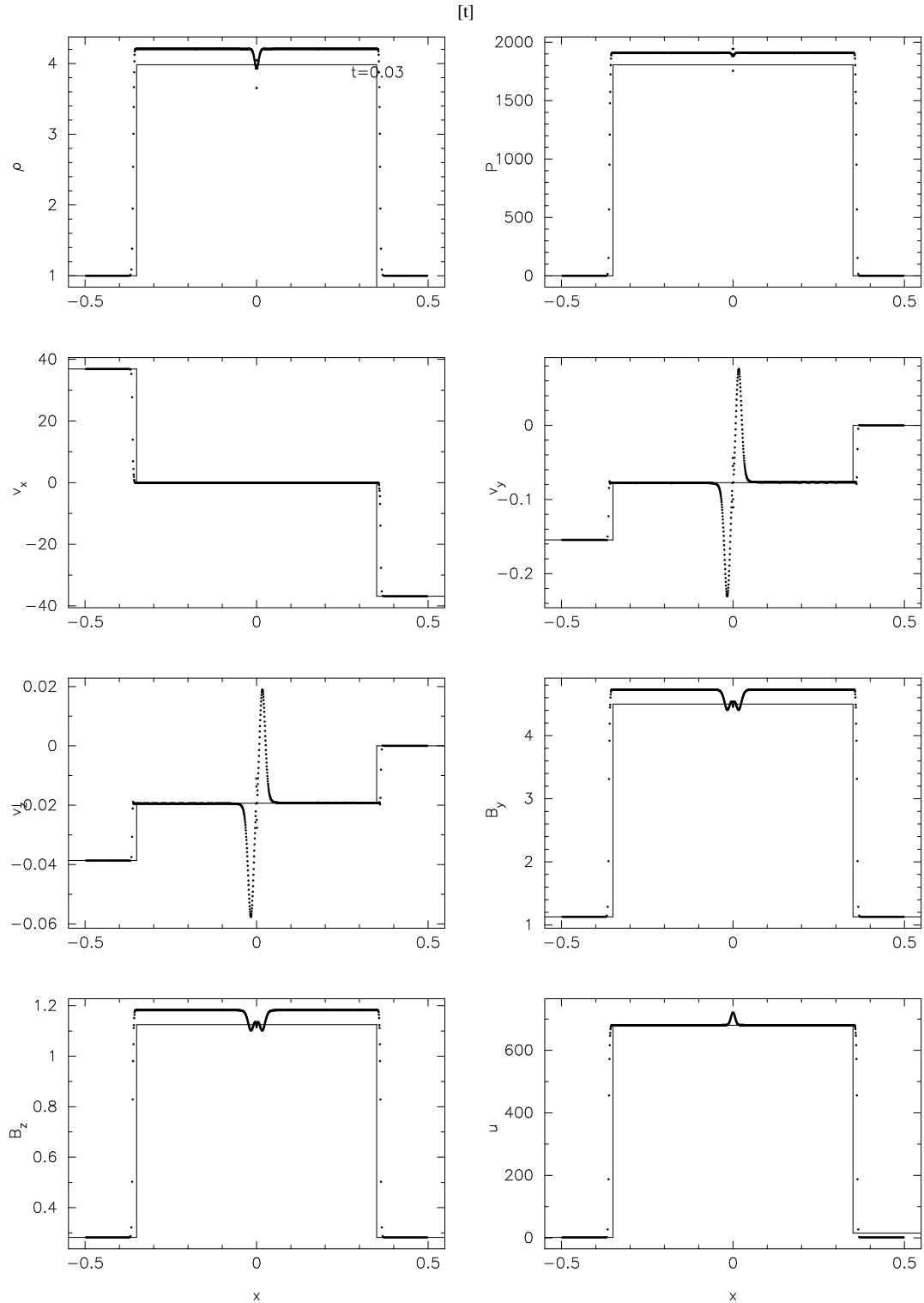}
\caption{Results of the MHD shock tube test with left state
$(\rho,P,v_x,v_y,v_z,B_y,B_z) = [1,1,36.87,-0.155,-0.0386,4/(4\pi)^{1/2},1/(4\pi)^{1/2}]$ and right state
$(\rho,P,v_x,v_y,v_z,B_y,B_z) = [1,1,-36.87,0,0,4/(4\pi)^{1/2},1/(4\pi)^{1/2}]$ with $B_x = 4.0/(4\pi)^{1/2}$ and $\gamma = 5/3$. Results
are shown at time $t=0.03$. This problem illustrates the formation of two
extremely strong fast shocks of Mach number 25.5 each.
Solid points indicate the position of the SPH particles whilst the exact
solution is given by the solid line. The artificial
dissipation switch is used with $K_{min} = 0.05$. The overshoots in density,
pressure and magnetic field are a result of our integration of the continuity
equation and neglect of terms relating to the gradient of the smoothing length.
(these terms are derived in paper II).}
\label{fig:balsara4}
\end{center}
\end{figure*}

 The fourth test (Figure \ref{fig:balsara5}) is similar to the previous version except that an isothermal
equation of state is used. The left state is given by
$(\rho,v_x,v_y,v_z,B_y,B_z) = [1.08,1.2,0.01,0.5,3.6/(4\pi)^{1/2},2/(4\pi)^{1/2}]$ and the right state
$(\rho,v_x,v_y,v_z,B_y,B_z)=[1,0,0,0,4/(4\pi)^{1/2},2/(4\pi)^{1/2}]$ with $B_x =
2/(4\pi)^{1/2}$ and an isothermal sound speed of unity. Results are shown in
Figure \ref{fig:balsara5} at time $t=0.2$ and compare very well with the
numerical results given in \citet{balsara98} (solid line).

 The fifth test shows the formation of two magnetosonic rarefactions. The left state is given by
$(\rho,P,v_x,v_y,B_y) = [1,1,-1,0,1]$ and the right state by
$(\rho,P,v_x,v_y,B_y)=[1,1,1,0,1]$ with $B_x = 0$ and $\gamma = 5/3$. Results
are shown in Figure \ref{fig:rarefaction} at time $t=0.1$ and compare extremely
well with the exact solution from \citet{rj95} (solid line). Outflow
boundary conditions are used such that the resolution varies from an initial
500 particles down to 402 particles at $t=0.1$ in the domain $x = [-0.5, 0.5]$.
The artificial dissipation switch is turned on with $K_{min} = 0.05$ although very little
dissipation occurs in this simulation since the artificial dissipation is only applied for particles
approaching each other. With unsmoothed initial conditions we therefore observe some oscillations behind the
rarefaction waves, which are removed in this case by smoothing
the initial discontinuity slightly. As noted in \citet{monaghan97} use of the
density summation also improves the results for this type of problem.

 The final test, taken from \citet{dw94} and \citet{balsara98}, illustrates the formation of
two fast shocks, each with Mach number 25.5.  The left state is given by
$(\rho,P,v_x,v_y,v_z,B_y,B_z) = [1,1,36.87,-0.155,-0.0386,4/(4\pi)^{1/2},1/(4\pi)^{1/2}]$ and the right state by
$(\rho,P,v_x,v_y,v_z,B_y,B_z) = [1,1,-36.87,0,0,4/(4\pi)^{1/2},1/(4\pi)^{1/2}]$ with $B_x = 4.0/(4\pi)^{1/2}$ and $\gamma = 5/3$. Results
are shown in Figure \ref{fig:balsara4} at time $t=0.03$. Inflow
boundary conditions are used such that the resolution varies from an initial
400 particles up to 1286 particles at $t=0.03$ in the domain $x = [-0.5, 0.5]$.
The artificial dissipation switch is turned on with $K_{min} = 0.05$.
 The results compare extremely well with the exact solution (solid line) given 
by \citet{dw94} and with the numerical solution given by \citet{dw94} and \citet{balsara98}, especially given the extreme nature of the problem. The
spikes in transverse velocity components are starting errors due to the fact
that for this problem we do not smooth the initially discontinuous velocity profiles in
any way. There is some advantage to integrating the total energy equation for
this type of problem since using the thermal energy equation produces a large
spike in thermal energy at the discontinuity and numerical noise behind the
shocks. The overshoots in density,
pressure and magnetic field are a result of our integration of the continuity
equation and neglect of terms relating to the gradient of the smoothing length.
These terms are derived and implemented in paper II and are shown to remove the errors seen
here.

\section{Summary}
We have shown how SPH equations for MHD can be formulated with the 
following features:

\begin{enumerate}
\item The equations use the continuum equations of \citet{janhunen00} and
\citet{dellar01} which are consistent even when the divergence of the magnetic field is 
non zero.  Consequently, even though non zero $\divB$ may be produced 
during the simulation,  it is treated consistently.  We find that 
insisting on consistency with fundamental principles is the key to 
deriving stable SPH equations.

\item The equations contain artificial dissipation. We require these 
dissipation terms to result in positive definite changes to the entropy 
and this places strong constraints on the form of the dissipation. Our 
equations guarantee that the resulting viscous and ohmic dissipation 
produces  positive changes in  the thermal energy .   In ensuring these 
equations are consistent with the fundamental requirement that the 
entropy should increase we are led to introduce a term in the induction 
equation which is analogous to the induction equation for non ideal MHD. The use
of the \citet{mm97} switch very effectively minimizes the effect of the
artificial dissipation away from shocks.

\item The SPMHD equations also incorporate a simple technique to prevent an 
instability due to the tension arising from the magnetic stress.  
\end{enumerate}

 The resulting equations, when implemented with a simple predictor 
corrector scheme, give good results for a wide range of shock tube 
problems.  While we have yet to apply our algorithm to problems in two 
and three dimensions the present results encourage us to believe that 
our SPMHD code will provide a secure basis for astrophysical MHD 
problems.

\appendix

\section{Artificial stress}
\label{sec:appendix1}

In this appendix we describe some of the details of the artificial stress
required to prevent clumping.  We diagonalise the magnetic part of the stress
tensor by rotating the coordinate system so that the $z$ axis lies along the
magnetic field. The magnetic field is then ${\bf B'} = (0,0,B)$ and the stress
tensor is  has non zero components $M'_{xx} = - B^2/(2\mu_0)$,   $M'_{yy} = - 
B^2/(2 \mu_0)$ , and  $M'_{zz} =   B^2/(2\mu_0)$. The sign of the first two is
associated with compression and the sign of the third is associated with
tension.   To remove the tension term at close range we add a term to $M'_{zz}$
so that it is negative when the particles approach .   The term we choose is $R
B^2$, where
\begin{equation}
R =  -\frac{\epsilon}{2 \mu_0} \left ( \frac{W_{ab} }{W(\Delta p)}   \right )^n
\end{equation}
where $\epsilon \sim 0.4$ and $n\sim 4$ and in the tests shown in this paper we
have $\epsilon = 0.8$ and $n = 5$. The precise values of these quantities
is not important. $W_{ab} $ is the kernel and $W(\Delta p)$ is $W$ evaluated at
the average particle spacing.  

Rotating back to the original coordinates we find that the term we subtracted is
equivalent to defining a new magnetic  stress
\begin{equation}
M'_{ij} =  M_{ij}  +  R B_i B_j.
\end{equation}

\section{Positivity of the Entropy change}
\label{sec:entropy}
In this appendix we demonstrate that the dissipation terms introduced in \S\ref{sec:dissipation} lead to a
positive definite increase in the entropy.

 The second law of thermodynamics shows that the change of entropy per unit mass $s_a$ of particle $a$ is given by
\begin{equation}
T_a \frac{ds_a}{dt} = \frac{du_a}{dt} - \frac{P_a}{\rho^2_a} \frac{d \rho_a}{dt},
\label{eq:2ndlawthermo}
\end{equation}
where $T_a$ is the temperature (absolute) of particle $a$.

From (\ref{eq:uthermterms}), (\ref{eq:vdiss}), and (\ref{eq:Bdiss}), and noting that the second term of
(\ref{eq:2ndlawthermo}), when expressed in SPH form, cancels the first term of 
(\ref{eq:uthermterms}), we find  that
\begin{eqnarray}
T_a \frac{ds_a}{dt} & = & \sum_b \frac{m_b Kv_{sig} }{\bar{\rho_{ab}} } \left ( - [{\bf v}_a \cdot {\bf j}
-{\bf v}_b \cdot {\bf j} ]^2 \frac{}{}\right. \\  &  & - \left. \frac{1}{\mu_0 \bar \rho_{ab} } \left [ {\bf B}_{ab}^2 - ({\bf B}_{ab}
\cdot {\bf j})^2 \right ] + u_a - u_b   \right )r_{ab} F_{ab}. \nonumber
\end{eqnarray}

Because $F_{ab} \le 0$ we can rewrite this equation as 

\begin{equation}
T_a \frac{ds_a}{dt} = Q_a + \sum_b \frac{m_b Kv_{sig} }{ \bar \rho_{ab}} (u_a - u_b )r_{ab} F_{ab},
\end{equation}
where $Q_a \ge 0$ is the contribution to the entropy change from the viscous and
ohmic dissipation which we have shown is positive definite. The second term is a
heat conduction term which can increase or decrease the thermal energy of a
particle but it must not result in a decrease of the entropy of the system. 
Note that if $u_a > u_b$ then the thermal conduction causes a decrease in the
thermal energy of particle $a$. This form of the conduction term that arises
here is similar to that derived by \citet{cm99} for SPH heat conduction  problems.

The change in the total entropy $S$ is then
\begin{eqnarray}
\frac{dS}{dt} & = & \sum_a m_a \frac{ds_a}{dt} \nonumber \\
& = & \sum_a \frac{m_a Q_a}{T_a} + \sum_a\sum_b \zeta_{ab} \frac{(u_a - u_b)}{T_a},
\label{eq:totentrop}
\end{eqnarray}

where 
\begin{equation}
\zeta_{ab} = \frac{m_b Kv_{sig} r_{ab} F_{ab}}{ \bar \rho_{ab}} \le 0.
\end{equation}

We can interchange the labels in the second term of (\ref{eq:totentrop}) and combine with the original form to write this term as 

\begin{equation}
\frac12 \sum_a\sum_b \zeta_{ab} (u_a - u_b)\left ( \frac{1}{T_a} - \frac{1}{T_b}   \right ).
\end{equation}
If, as is normally the case, $u$ is a monotonically increasing function of $T$ then, if $T_a > T_b$ we have $u_a>u_b$ and 
\begin{equation}
(u_a - u_b)\left (   \frac{1}{T_a} - \frac{1}{T_b} \right ) \le 0
\end{equation}
so that the second term on the right hand side of (\ref{eq:totentrop}) is
positive.  The change in the entropy due to thermal conduction is therefore positive.  

\section*{Acknowledgements} 
DJP acknowledges the support of the Association of Commonwealth Universities and the Cambridge
Commonwealth Trust. He is supported by a Commonwealth Scholarship and Fellowship Plan. DJP
would also like to acknowledge useful discussions with Prof. J.E. Pringle.

\bibliography{/home/dprice/bibtex/sph,/home/dprice/bibtex/mhd}

\label{lastpage}
\enddocument